\documentclass[aps,superscriptaddress,prc,twocolumn,nofootinbib]{revtex4}

\usepackage{graphicx}
\usepackage{epstopdf}
\usepackage{subfigure}
\usepackage{epsfig}
\usepackage{amsmath,amssymb,amsfonts}
\usepackage{color}
\usepackage[utf8]{inputenc}
\allowdisplaybreaks
\usepackage[bookmarks,
                bookmarksopen = true,
                bookmarksnumbered = true,
                linktocpage,
                colorlinks = true,
                linkcolor = blue,
                urlcolor  = blue,
                citecolor = blue,
                anchorcolor = green,
                hyperindex = true,
                hyperfigures]
                {hyperref}

\begin{document}

\title{Perturbative and non-perturbative interactions between heavy quarks and quark-gluon plasma within a unified approach}

\author{Wen-Jing Xing}
\affiliation{Institute of Particle Physics and Key Laboratory of Quark and Lepton Physics (MOE), Central China Normal University, Wuhan, 430079, China}

\author{Guang-You Qin}
\email{guangyou.qin@mail.ccnu.edu.cn}
\affiliation{Institute of Particle Physics and Key Laboratory of Quark and Lepton Physics (MOE), Central China Normal University, Wuhan, 430079, China}

\author{Shanshan Cao}
\email{shanshan.cao@sdu.edu.cn}
\affiliation{Institute of Frontier and Interdisciplinary Science, Shandong University, Qingdao, Shandong 266237, China}

\date{\today}


\begin{abstract}

While perturbative QCD is sufficient for understanding the color, mass and energy dependences of parton energy loss and jet quenching at large transverse momentum in heavy-ion collisions, a simultaneous description of heavy flavor nuclear modification factor $R_\mathrm{AA}$ and elliptic flow coefficient $v_2$ at low and intermediate $p_\mathrm{T}$ still remains a challenge due to the effects from non-perturbative interactions. In this work, we extend the linear Boltzmann transport model by implementing a generalized Cornell-type potential that incorporates both short-range Yukawa interaction and long-range color confining interaction between heavy quarks and the QGP medium. Combining our new approach for heavy-quark-QGP interaction with a (3+1)-dimensional hydrodynamic model CLVisc for the QGP evolution and a hybrid fragmentation-coalescence model for heavy quark hadronization, we obtain a satisfactory description of heavy meson $R_\mathrm{AA}$ and $v_2$ from low to intermediate to high $p_\mathrm{T}$ observed at both RHIC and the LHC.
By model-data-comparison, we extract for the first time the in-medium heavy quark potential from open heavy flavor measurements; the result is in agreement with the lattice QCD calculation.
Our study indicates that while jet quenching at high $p_\mathrm{T}$ is dominated by perturbative QCD interactions, non-perturbative interactions are indispensible for understanding heavy flavor quenching and flow at low and intermediate $p_\mathrm{T}$.

\end{abstract}

\maketitle


\section{Introduction}
\label{sec:Introduction}

Heavy quarks serve as a clean probe of the quark-gluon plasma (QGP) matter produced in high-energy nuclear collisions, as they are primarily produced from the early-state hard scatterings and encode the information of the entire evolution history of the QGP fireballs. At low transverse momentum ($p_\mathrm{T}$), heavy quarks can be utilized to study the diffusion and thermalization processes inside the QGP~\cite{Moore:2004tg,Cao:2011et}. At intermediate $p_\mathrm{T}$, heavy quarks are excellent tracers of the color neutralization process when they convert into heavy flavor hadrons~\cite{Plumari:2017ntm,Song:2018tpv,He:2019vgs,Cho:2019lxb,Cao:2019iqs}. At high $p_\mathrm{T}$, heavy quarks provide a unique probe for investigating the flavor and mass dependences of parton energy loss and jet quenching~\cite{Zhang:2003wk,Djordjevic:2013pba,Zhang:2018nie,Du:2018yuf,Qin:2015srf,Cao:2020wlm,Xing:2019xae}. Phenomenologically, tremendous efforts have been devoted to simulating heavy quark dynamics in relativistic heavy-ion collisions using various model implementations ~\cite{Gossiaux:2006yu,Qin:2009gw,Gossiaux:2010yx,Das:2010tj,He:2011qa,Fochler:2013epa,Das:2015ana,Song:2015ykw,Cao:2017hhk,Xu:2017obm,Ke:2018tsh,Xing:2019xae,Li:2019wri,Liu:2021dpm,Rapp:2018qla,Cao:2018ews,Xu:2018gux,Katz:2019fkc}. Experimentally, a vast amount of data have been revealed at the Relativistic Heavy-Ion Collider (RHIC) and the Large Hadron Collider (LHC) on the nuclear modification factor ($R_\mathrm{AA}$) and collective flow coefficients ($v_1$, $v_2$ and $v_3$) of heavy flavor hadrons and their decayed leptons~\cite{PHENIX:2006iih,STAR:2014wif,STAR:2017kkh,CMS:2017qjw,CMS:2017vhp,ALICE:2017pbx,ALICE:2018lyv,STAR:2018zdy,ATLAS:2018ofq,STAR:2019clv,ALICE:2019sgg,ATLAS:2020yxw}.

While perturbative calculation has been successful in understanding heavy and light flavor jet quenching at high $p_\mathrm{T}$~\cite{Xing:2019xae,Liu:2021izt}, it fails at low $p_\mathrm{T}$~\cite{Caron-Huot:2008dyw}. Effects of non-perturbative interactions become crucial in describing heavy flavor dynamics from low to intermediate $p_\mathrm{T}$ region ($p_\mathrm{T} \lesssim 8-10$~GeV). This involves the strong coupling between heavy quarks and QGP, and the hadronization process of heavy quarks. To probe heavy quark hadronization, one important observation is the enhancement of the $D_s / D^0$ and $\Lambda_c / D^0$ ratios in heavy-ion collisions compared to proton-proton collisions at both RHIC~\citep{Zhou:2017ikn,STAR:2019ank} and the LHC~\citep{ALICE:2018lyv,ALICE:2018hbc}. These features have been attributed to the quark coalescence mechanism in Refs.~\cite{Plumari:2017ntm,Song:2018tpv,He:2019vgs,Cho:2019lxb,Cao:2019iqs}.
In particular, Ref.~\cite{Cao:2019iqs} shows that the interplay between the QGP flow and the charm quark $p_\mathrm{T}$ spectrum is essential for understanding the larger $\Lambda_c / D^0$ ratio observed at RHIC than at the LHC. Regarding the interaction of heavy quarks with QGP, the large elliptic flow coefficient ($v_2$) observed for heavy mesons and their decayed leptons from low to intermediate $p_\mathrm{T}$ still challenges most perturbative QCD based model calculations. In order to describe heavy flavor $R_\mathrm{AA}$ and $v_2$ data, these models usually apply a $K$ factor ~\cite{Nahrgang:2013saa,Cao:2016gvr} or invoke various assumptions on the running coupling $\alpha_\mathrm{s}$~\cite{Gossiaux:2010yx,Das:2015ana} to enhance heavy-quark-QGP interactions at low $p_\mathrm{T}$.

A more delicate treatment of the non-perturbative interactions at low momentum is to apply the in-medium color force (potential) between heavy quark and antiquark, which can be characterized by a short-range Yukawa interaction and a long-range color confining interaction. For instance, the interaction potential extracted from the finite-temperature lattice QCD calculation has been used in a $T$-matrix approach~\citep{vanHees:2007me,Riek:2010fk,Liu:2018syc,He:2019vgs,He:2011qa}. With the inclusion of non-perturbative effects from the color confining interaction, the resulting drag coefficient increases with decreasing medium temperature, which is contrary to the trend predicted by perturbative QCD calculations. As shown by Ref.~\citep{Das:2015ana}, such temperature dependence of the drag coefficient near the transition temperature ($T_\mathrm{c}$) is essential for a simultaneous description of the heavy flavor $R_\mathrm{AA}$ and $v_2$ at low to intermediate $p_\mathrm{T}$. In the pioneer studies above, the confining interaction between heavy-light quark scattering is implemented in the Langevin transport model through the heavy quark diffusion coefficients. In our present work, we incorporate the confining interaction into the linear Boltzmann transport (LBT) model~\cite{Cao:2016gvr,Cao:2017hhk} which can simultaneously describe elastic and inelastic scatterings of heavy and light flavor partons inside the QGP.
We improve the LBT model by re-evaluating the heavy quark scattering rates including both perturbative Yukawa and non-perturbative color confining interactions between heavy quarks and thermal partons inside the QGP. By coupling this improved LBT model to the (3+1)-D viscous hydrodynamic model CLVisc~\cite{Pang:2012he,Pang:2018zzo,Wu:2018cpc,Wu:2021fjf} for the QGP evolution and the hybrid fragmentation-coalescence model for heavy quark hadronization, we obtain a satisfactory description of the $D$ meson $R_\mathrm{AA}$ and $v_2$ observed at RHIC and the LHC from low to intermediate to high $p_\mathrm{T}$. From the model-to-data comparison, we extract for the first time the in-medium heavy quark potential from open heavy flavor observables. Our extracted potential is consistent with the lattice QCD data.
We also compute the momentum and temperature dependences of the jet quenching parameter ($\hat{q}$) and the spatial diffusion coefficient of heavy quarks ($D_\mathrm{s}$).

This paper is organized as follows. In Sec.~\ref{sec:new_M2}, we extend the perturbative calculation of heavy-light parton scattering matrix elements to include contributions from both short-range Yukawa and long-range color confining interactions. In Sec.~\ref{sec:modified_LBT}, we apply our improved approach to a realistic simulation of heavy quark production, medium modification and hadronization in relativistic heavy-ion collisions. Numerical results, including the heavy meson $R_\mathrm{AA}$ and $v_2$, the extracted in-medium heavy quark potential and transport coefficients, are presented in Sec.~\ref{sec:results}. We summarize in Sec.~\ref{sec:summary}.

\section{Perturbative and non-perturbative interactions between heavy quarks and QGP}
\label{sec:new_M2}

As shown by Refs.~\cite{Gossiaux:2008jv,He:2011qa,Xu:2014tda,Das:2015ana}, perturbative description of elastic scatterings between heavy quarks and the QGP is not sufficient to understand the heavy meson suppression and elliptic flow observed at RHIC and the LHC, especially in the low to intermediate $p_\mathrm{T}$ regime.
Meanwhile, lattice QCD calculations~\cite{Petreczky:2004pz,Kaczmarek:2005ui} indicate the non-perturbative effects from the long-range confining interaction part of the heavy quark pair ($Q\bar{Q}$) potential can survive up to a temperature of at least twice of the value of phase transition temperature.
In Ref.~\cite{Liu:2018syc}, the color potential is first obtained by fitting to the lattice QCD data on heavy quark free energy, quarkonium correlators and the QGP equation of state.
Then the obtained potential is applied to evaluate the heavy quark transport coefficients for simulating the dynamical evolution of heavy quarks in the Langevin transport approach~\cite{He:2011qa}.
In this work, we include the non-perturbative interaction between heavy quarks and thermal partons in the LBT model, which treats elastic and inelastic energy loss of heavy quarks in the QGP on the same footing.

We start with a parametrized Cornell-type potential inspired from Ref.~\cite{Liu:2017qah} for the heavy-quark-QGP scattering. It includes both short-range Yukawa potential and long-range color confining potential (or ``string term") as follows:
\begin{eqnarray}
  \label{eq:Cornell_poten1}
 V(r) = V_\mathrm{Y}(r) + V_\mathrm{S}(r) = -\frac{4}{3} \alpha_\mathrm{s} \frac{e^{-m_d r}}{r} - \frac{\sigma e^{-m_s r}}{m_s}.
\end{eqnarray}
In the above equation, $\alpha_\mathrm{s}$ and $\sigma$ are the coupling strengths for Yukawa and string terms.
The temperature dependent screening masses are taken as $m_d = a + bT$ and $m_s = \sqrt{a_s + b_sT}$.
One can see that the above potential is similar to the heavy quark pair potential $V_{\mathrm{Q \bar{Q}}}$ in vacuum~\cite{Koma:2006si} except that in a thermal medium, additional scales (such as the medium temperature) enter both Yukawa and string parts of the potential.
In our current work, instead of fixing parameters $\alpha_\mathrm{s},\sigma, a,b, a_s, b_s$ from the lattice QCD data, we leave them as model parameters, which will be determined by comparing our model calculations of the heavy meson quenching and flow observables to the experimental data. This allows us to extract the in-medium heavy quark potential from open heavy flavor measurements in heavy-ion collisions.

Using the Fourier transformation, the above Cornell-type potential can be expressed in the momentum space as follows:
\begin{eqnarray}
  \label{eq:HQ_potential}
  V(\vec{q}) = - \frac{4\pi \alpha_\mathrm{s} C_F}{m_d^2+|\vec{q}|^2} - \frac{8\pi \sigma}{(m_s^2+|\vec{q}|^2)^2},
\end{eqnarray}
where $\vec{q}$ denotes the momentum transfer between heavy quarks and thermal partons from the QGP medium.

To implement the above potential in calculating the scattering rates of heavy quarks in the QGP, we first consider a two body scattering process, $Qq \rightarrow Qq$, in which a gluon with momentum $\vec{q}$ is exchanged. Without quantizing the strong interaction gauge field, we treat the in-medium Cornell-type potential in Eq.~(\ref{eq:HQ_potential}) as the effective gluon propagator (field). Following Ref.~\cite{Riek:2010fk}, we assume a scalar interaction vertex for the string term, while keeping a vector interaction vertex for the Yukawa term to be consistent with the leading-order perturbative QCD calculation of two body scattering. The scattering amplitude is then expressed as
\begin{align}
  \label{eq:Matrix_cq}
 i \mathcal{M}  =\,& \mathcal{M_\mathrm{Y}} + \mathcal{M_\mathrm{S}}\nonumber
 \\=\,& \overline{u}(p') \gamma^{\mu} \overline{u}(p) V_\mathrm{Y}(\vec{q}) \overline{u}(k') \gamma^{\nu} \overline{u}(k)\nonumber
 \\&+ \overline{u}(p') \overline{u}(p) V_\mathrm{S}(\vec{q}) \overline{u}(k') \overline{u}(k),
\end{align}
in which $\mathcal{M_\mathrm{Y}}$ and $\mathcal{M_\mathrm{S}}$ represent the Yukawa and string contributions to the scattering amplitude, respectively. The color information of interaction vertices has already been included in the heavy quark potential. By setting the momentum exchange as $|\vec{q}|^2 = -t$ in the center of mass frame, we obtain the amplitude squared for the $Qq \rightarrow Qq$ scattering process as follows,
\begin{align}
  \label{eq:M2_cq}
|\mathcal{M}_\mathrm{Qq}|^2 &= \frac{64\pi^2 \alpha_\mathrm{s}^2}{9} \frac{(s-m_Q^2)^2+(m_Q^2-u)^2+2m_Q^2 t}{(t-m_d^2)^2}\nonumber\\
 &+ \frac{(8\pi \sigma)^2}{N_c^2 -1} \frac{t^2-4m_Q^2 t}{(t-m_s^2)^4},
\end{align}
with $s$, $t$, $u$ being the Mandelstam variables. In the above equation, the sum of final-state and the average over initial-state spin degrees of freedom have been included. The first term corresponds to the perturbative-based Yukawa scattering matrix element, while the second term denotes the string contribution. To reproduce the color factor $C_F^2/(N_c^2-1)$ of the well-established leading-order perturbative QCD result of heavy-light quark scattering~\cite{Combridge:1978kx} in the first term, an extra color-average factor $1/(N_c^2-1)$ for the gluon field has been introduced in obtaining Eq.~(\ref{eq:M2_cq}).
Due to different types of the interaction vertices, there is no interference term between the Yukawa scattering amplitude ($\mathcal{M}_\mathrm{Y}$) and the string scattering amplitude ($\mathcal{M}_\mathrm{S}$).
Note that the $|\mathcal{M}_\mathrm{Y}|^2$ term is the same as the perturbative result in Ref.~\cite{Combridge:1978kx} except the Debye screening mass squared $m_d^2$ is introduced in the gluon propagator.

It is well known that while the $Qq \rightarrow Qq$ scattering process has only the $t$-channel contribution, the $Qg \rightarrow Qg$ process has contributions from $s$, $t$ and $u$-channels altogether.
Since the in-medium potential represents the effective gluon propagator, we only need to introduce the string term for the $t$-channel scattering.
As for the Yukawa contribution, one only needs to replace $t$ by $t-m_d^2$ in the denominators of the leading-order perturbative QCD result.
Putting together, the final amplitude squared for the $Qg \rightarrow Qg$ process is given by:
\begin{align}
  \label{eq:M2_cg}
  |\mathcal{M}&_{Qg}|^2 = \nonumber\\
  & \frac{64\pi^2 \alpha_\mathrm{s}^2}{9} \frac{(s-m_Q^2)(m_Q^2-u)+2m_Q^2 (s+m_Q^2)}{(s-m_Q^2)^2}\nonumber
  \\+\, & \frac{64\pi^2 \alpha_\mathrm{s}^2}{9} \frac{(s-m_Q^2)(m_Q^2-u)+2m_Q^2 (u+m_Q^2)}{(u-m_Q^2)^2}\nonumber
  \\+\, & 8\pi^2 \alpha_\mathrm{s}^2 \frac{5m_Q^4 + 3m_Q^2t -10m_Q^2u + 4t^2 + 5tu + 5u^2}{(t-m_d^2)^2}\nonumber
  \\+\, & 8\pi^2 \alpha_\mathrm{s}^2 \frac{(m_Q^2-s)(m_Q^2-u)}{(t-m_d^2)^2}\nonumber
  \\+\, & 16\pi^2 \alpha_\mathrm{s}^2 \frac{3m_Q^4 -3m_Q^2s - m_Q^2u + s^2}{(s-m_Q^2)(t-m_d^2)}\nonumber
  \\+\, & \frac{16\pi^2 \alpha_\mathrm{s}^2}{9} \frac{m_Q^2 (4m_Q^2 - t)}{(s - m_Q^2)(m_Q^2 - u)}\nonumber
  \\+\, & 16\pi^2 \alpha_\mathrm{s}^2 \frac{3m_Q^4 - m_Q^2s -3m_Q^2u +u^2}{(t - m_d^2)(u - m_Q^2)}\nonumber
  \\+\, & \frac{C_A}{C_F} \frac{(8\pi \sigma)^2}{N_c^2 -1} \frac{t^2-4m_Q^2 t}{(t-m_s^2)^4}.
\end{align}
Similar to Eq.~(\ref{eq:M2_cq}), a $1/(N_c^2-1)$ factor has been introduced.
Note that the additional $C_A/C_F$ factor in the string term is to account for different color factors between the $Qg \rightarrow Qg$ process and the $Qq \rightarrow Qq$ process.
We can see that the matrix element squared contains a perturbative part ($|\mathcal{M}_\mathrm{Y}|^2$) and a non-perturbative part ($|\mathcal{M}_\mathrm{S}|^2$).

\section{Heavy quark transport in QGP}
\label{sec:modified_LBT}

Now we implement the perturbative and non-perturbatvie interactions in the linear Boltzmann transport (LBT) model~\cite{He:2015pra,Cao:2017hhk,Cao:2016gvr} for simulating heavy quark evolution inside a realistic QGP medium.
In the LBT model, the spacetime distribution of heavy quarks (denoted as ``$a$") is evolved according to the Boltzmann equation as
\begin{eqnarray}
  \label{eq:boltzmann1}
  p_a \cdot\partial f_a(x,p)=E_a (\mathcal{C}_\mathrm{el}+\mathcal{C}_\mathrm{inel}),
\end{eqnarray}
where $\mathcal{C}_\mathrm{el}$ and $\mathcal{C}_\mathrm{inel}$ are the collision integrals for elastic and inelastic scatterings, respectively.

For elastic scattering between heavy quarks and medium constituents, we apply the matrix elements $|\mathcal{M}_{ab \rightarrow cd}|^2$ derived in the previous section to calculate the scattering rate as follows:
\begin{align}
  \label{eq:gamma_el}
   \Gamma_{ab \rightarrow cd} & (\vec{p}_a, T) = \frac{\gamma_2}{2E_a}\int \frac{d^3 p_b}{(2\pi)^3 2E_b} \frac{d^3 p_c}{(2\pi)^3 2E_c} \frac{d^3 p_d}{(2\pi)^3 2E_d}\nonumber
  \\ \times\,& f_b (\vec{p}_b, T) [1 \pm f_c (\vec{p}_c, T)] [1 \pm f_d (\vec{p}_d, T)]\nonumber
  \\ \times\, &\theta (s - (m_a + \mu_d)^2) \nonumber
  \\ \times\,& (2\pi)^4 \delta^{(4)}(p_a + p_b - p_c -p_d) |\mathcal{M}_{ab \rightarrow cd}|^2.
\end{align}
In this work, we assume the medium partons are massless, and take $M_c = 1.27$~GeV for charm quarks.
In the above equation, $f_b$ and $f_d$ are thermal distributions of the medium partons, and the $1-f_c$ factor is neglected for heavy quarks considering their dilute distribution inside the QGP when the medium temperature is small compared to $M_c$.
To account for the thermal mass effect of the medium partons, a $\theta$-function is introduced, with $\mu_d$ being the same temperature-dependent Debye mass used in the Yukawa potential.
Following our earlier work~\cite{Cao:2017hhk,Xing:2019xae}, we use different values of the coupling strength $\alpha_\mathrm{s}$ for different interaction vertices. For the vertex connecting the jet parton (heavy quark), an energy and temperature dependent running coupling $\alpha_\mathrm{s} = 4\pi / [9 \mathrm{ln}(2ET/\Lambda^2)]$ with $\Lambda=0.2$~GeV is applied. For the vertex connecting to the medium partons inside the QGP, we take the same $\alpha_\mathrm{s}$ value as used in the Yukawa potential.

With the above setups, the total elastic scattering rate $\Gamma_\mathrm{el}^a$ of a given heavy quark inside the QGP is obtained by summing over all possible scattering channels: $\Gamma_\mathrm{el}^a = \Sigma_i \Gamma_{el}^i$. For a given time step $\Delta t$, the elastic scattering probability for a heavy quark is given by $P_\mathrm{el}^a = 1-e^{- \Gamma_\mathrm{el}^a \Delta t}$.

For the inelastic process, the scattering rate is given by the average number of emitted gluons per unit time:
\begin{eqnarray}
  \label{eq:gamma_inel}
  \Gamma_\mathrm{inel}^a (E_a, T, t) = \int dxdl_{\bot}^2 \frac{dN_g^a}{dxdl_\bot ^2 dt},
\end{eqnarray}
where the time (length) dependent gluon spectrum is taken from the higher-twist energy loss formalism~\cite{Wang:2001ifa, Zhang:2003wk, Majumder:2009ge},
\begin{eqnarray}
  \label{eq:gluon_spectrum}
  \frac{dN_g^a}{dxdl_\bot ^2 dt} = \frac{2C_A \alpha_\mathrm{s} P_a(x) l_{\bot}^4 \hat{q}_a}{\pi (l_{\bot}^2 + x^2 m_a^2)^4} \sin^2 \left(\frac{t-t_i}{2\tau_f}\right).
\end{eqnarray}
Here, $x$ and $l_{\perp}$ denote the energy fraction and the transverse momentum of the radiated gluon with respect to its parent parton, $P_a(x)$ is the vacuum splitting function, $t-t_i$ gives the accumulative time since the previous emission (at $t_i$), and $\tau_f = 2Ex(1-x)/(l_{\bot}^2+x^2m_a^2)$ is the gluon formation time, with $E_a$ and $m_a$ being the jet parton energy and mass, respectively.
The running coupling for the jet parton splitting vertex is taken as $\alpha_\mathrm{s} = 4\pi / [9 \mathrm{ln}(2ET/\Lambda^2)]$.
In addition, $\hat{q}_a$ is the jet transport coefficient that quantifies the transverse momentum broadening of the jet parton per unit time due to elastic scattering, which can be obtained by evaluating Eq.~(\ref{eq:gamma_el}) with $k^2_\perp=[\vec{p}_c-(\vec{p}_c\cdot \vec{p}_a)\vec{p}_a/|\vec{p}_a|]^2$ as a weight inside the integral. With Eq.~(\ref{eq:gamma_inel}), the probability of inelastic scattering during a time step $\Delta t$ is then given by $P_\mathrm{inel}^a = 1-e^{-\Gamma_\mathrm{inel}^a \Delta t}$.

To simultaneously include elastic and inelastic processes, one may write the total scattering rate as $\Gamma_\mathrm{tot}=\Gamma_\mathrm{el}+\Gamma_\mathrm{inel}$. This leads to,
\begin{eqnarray}
  \label{eq:prob_total}
  P_\mathrm{tot}^a = 1 - e^{-\Gamma_\mathrm{tot} \Delta t} = P_\mathrm{el}^a + P_\mathrm{inel}^a - P_\mathrm{el}^a P_\mathrm{inel}^a.
\end{eqnarray}
The total interaction probability can be understood as the sum of the pure elastic scattering probability $P_\mathrm{el}^a (1-P_\mathrm{inel}^a)$ and the inelastic scattering probability $P_\mathrm{inel}^a$.

The above improved LBT model is coupled to the (3+1)-D viscous hydrodynamic model CLVisc~\cite{Pang:2012he,Pang:2018zzo,Wu:2018cpc,Wu:2021fjf} for the evolution of heavy quarks inside a realistic QGP medium. The CLVisc calculation provides the space-time profile, including the local temperature and flow velocity, of the QGP fireball produced in heavy-ion collisions at RHIC and the LHC.
Using the medium information, each heavy quark at a given time step is first boosted to the local rest frame of the expanding medium.
If a scattering process happens, one updates the heavy quark's energy and momentum according to the procedures described above.
Then the heavy quark is boosted back to the global frame of heavy-ion collisions and stream freely to the next location for possible interaction in the next time step.

In this work, the Glauber model is used for the initial spatial distributions of both heavy quarks and the energy density of the QGP medium. The momentum space of heavy quarks is initialized with the fixed-order-next-to-leading-log (FONLL) calculation~\cite{Cacciari:2001td,Cacciari:2012ny,Cacciari:2015fta} convoluted with the CT14NLO~\cite{Dulat:2015mca} parton distribution function. The EPPS16~\cite{Eskola:2016oht} parametrization at the next-to-leading-order is applied for the nuclear shadowing effect in nucleus-nucleus collisions. Heavy quarks are assumed to stream freely before the hydrodynamic expansion of the medium starts (at $\tau_0=0.6$~fm). After that, heavy quarks interact with the hydrodynamic medium until they exit the color-deconfined region of the nuclear matter. On the hypersurface of $T_\mathrm{c}=165$~MeV, heavy quarks are converted into heavy flavor hadrons through a hybrid fragmentation-coalescence model~\cite{Cao:2019iqs}, which takes into account both $s$ and $p$ wave states of the produced hadrons and can successfully describe the heavy flavor hadron chemistry observed at RHIC and the LHC. In the present study, smooth hydrodynamic profiles with specific shear viscosity $\eta/s=0.08$ (for RHIC) and $0.16$ (for LHC) are used. Effects of event-by-event fluctuations on heavy flavor observables are found small in our earlier work~\cite{Cao:2014fna,Cao:2017umt}. Other model uncertainties, such as the initial spectrum of heavy quarks and their evolution during the pre-equilibrium stage of the medium, have been discussed in detail in Ref.~\cite{Li:2020kax}.

\section{Heavy meson observables and extraction of the color potential}
\label{sec:results}

Using the improved LBT model that includes both perturbative (Yukawa) and non-perturbative (confining/string) interactions between heavy quarks and the QGP medium, we calculate the nuclear modification factor $R_\mathrm{AA}$ and the elliptic flow coefficient $v_2$ of $D$ mesons in this section. By comparing our theoretical results to the experimental data, the model parameters are extracted as: $\alpha_\mathrm{s}=0.27$, $\sigma=0.45$~GeV$^2$, $m_d=2T+0.2$~GeV, and $m_s=\sqrt{0.1~\mathrm{GeV}\times T}$. With these parameters, we calculate the heavy quark in-medium potential and transport coefficients $\hat{q}$ and $D_\mathrm{s}$.

\begin{figure}[tbp]
\centering
\includegraphics[width=0.97\linewidth]{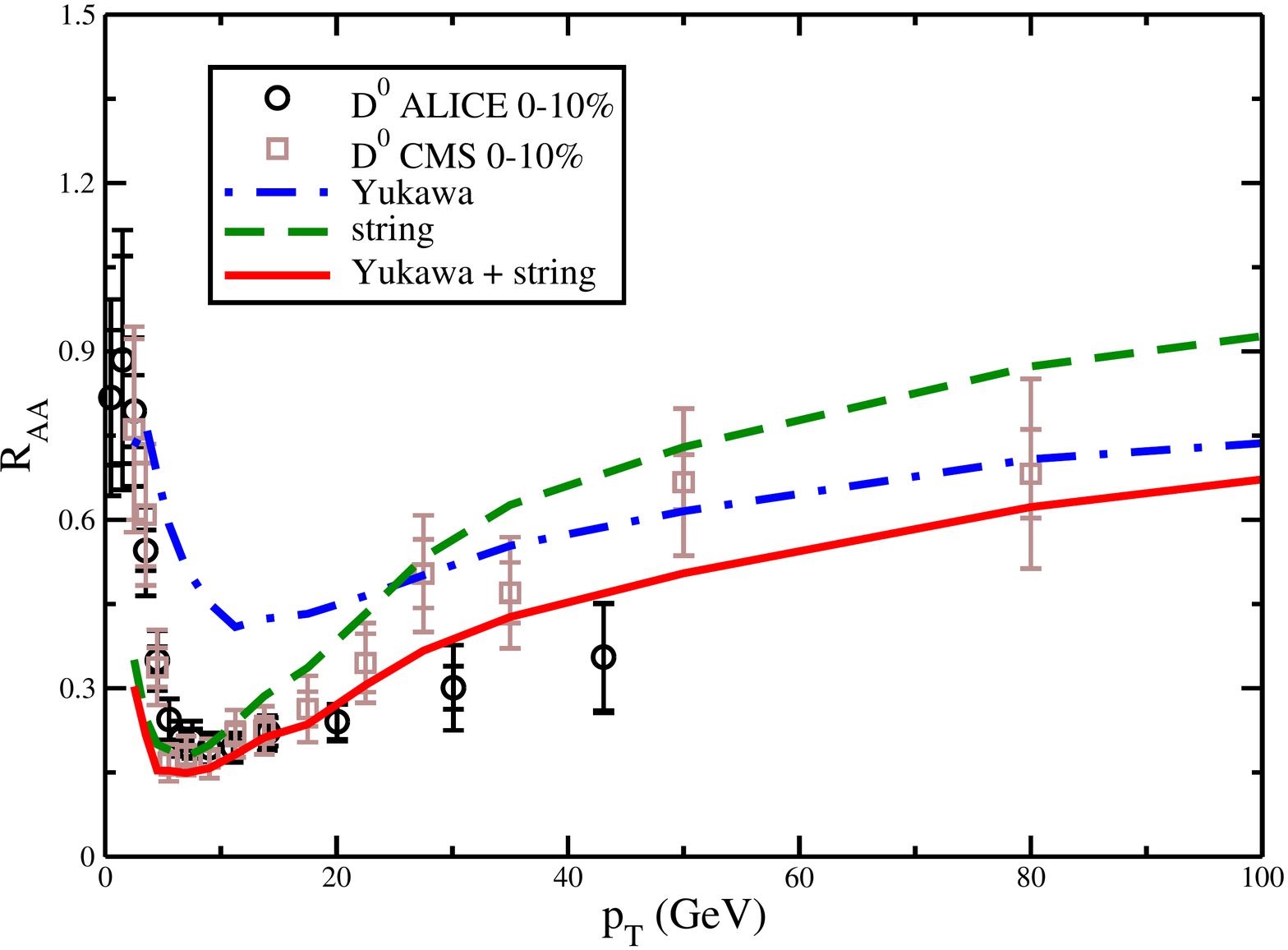}
\includegraphics[width=0.97\linewidth]{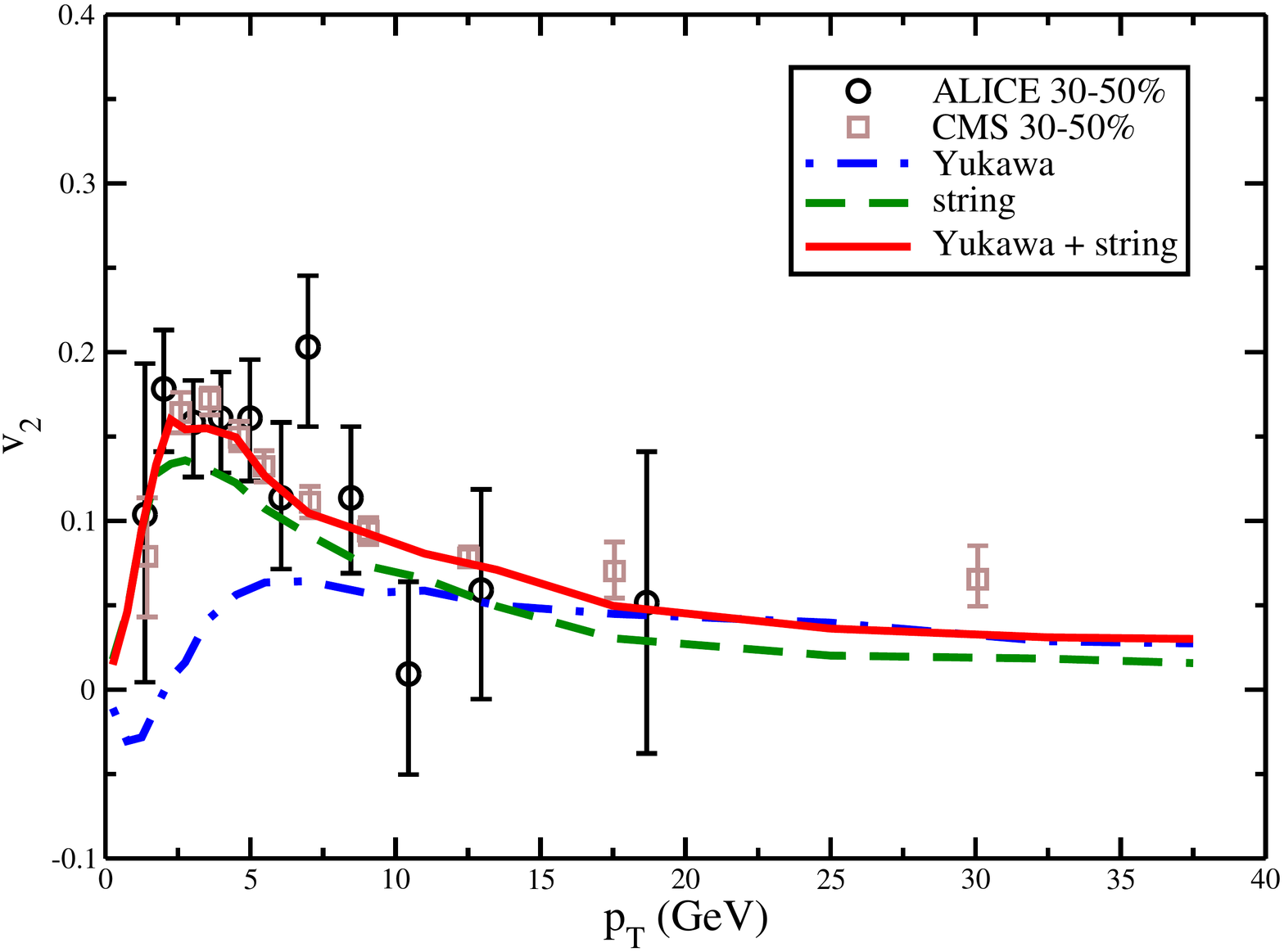}
\caption{(Color online) The $D$ meson $R_\mathrm{AA}$ and $v_2$ in Pb+Pb collisions at 5.02~ATeV, compared to the CMS and ALICE data~\cite{CMS:2017qjw,ALICE:2018lyv,CMS:2017vhp,ALICE:2017pbx}.}
\label{Fig:LHC-Dmeson_RAA_v2}
\end{figure}

\begin{figure}[tbp]
\includegraphics[width=0.97\linewidth]{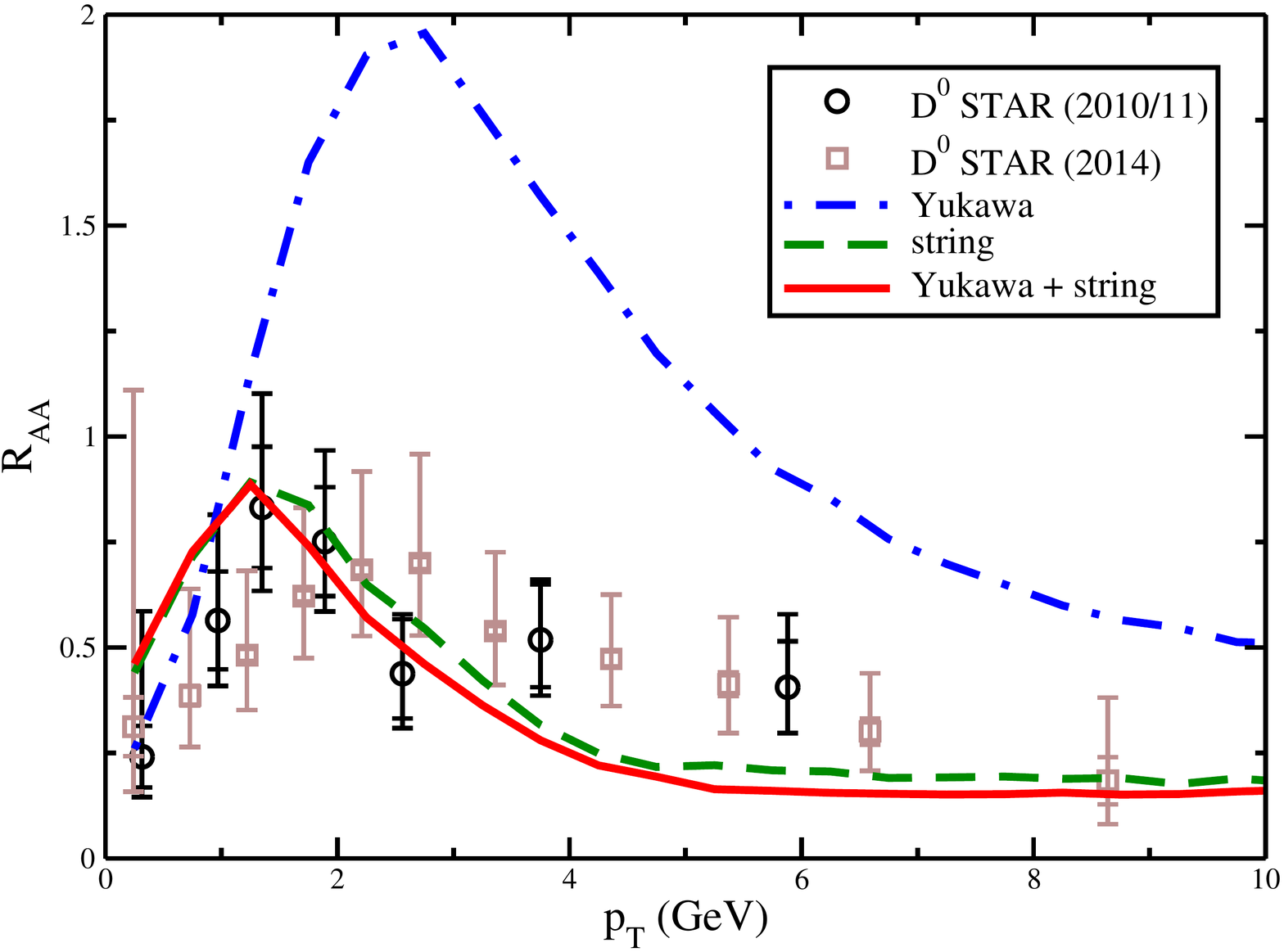}
\includegraphics[width=0.97\linewidth]{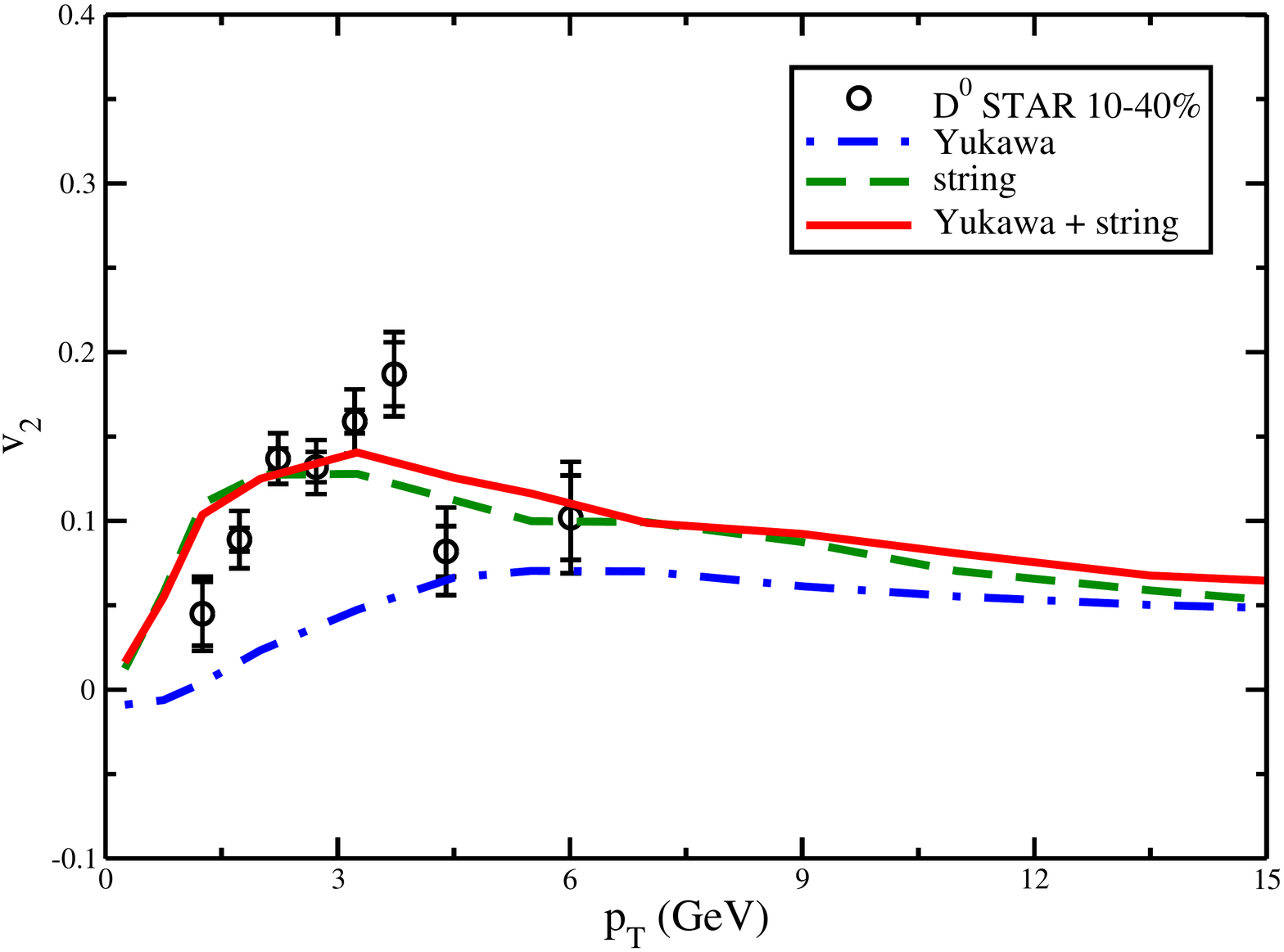}
\caption{(Color online) The $D$ meson $R_\mathrm{AA}$ and $v_2$ in Au+Au collisions at 200~AGeV, compared to the STAR data~\cite{STAR:2014wif,STAR:2018zdy,STAR:2017kkh}.}
	\label{Fig:RHIC-Dmeson_RAA_v2}
\end{figure}

\begin{figure}[tbp]
\centering
\includegraphics[width=0.97\linewidth]{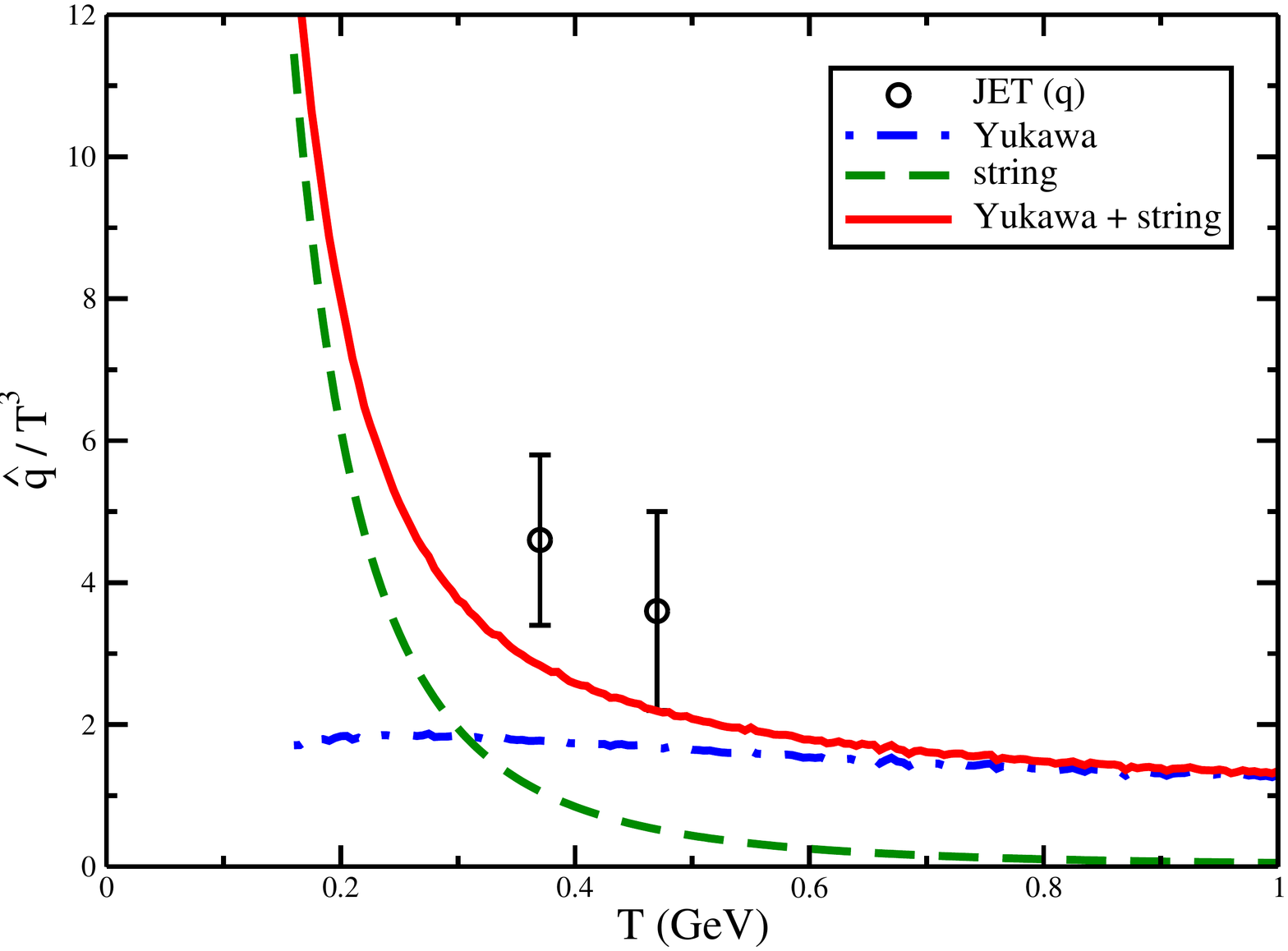}
\includegraphics[width=0.97\linewidth]{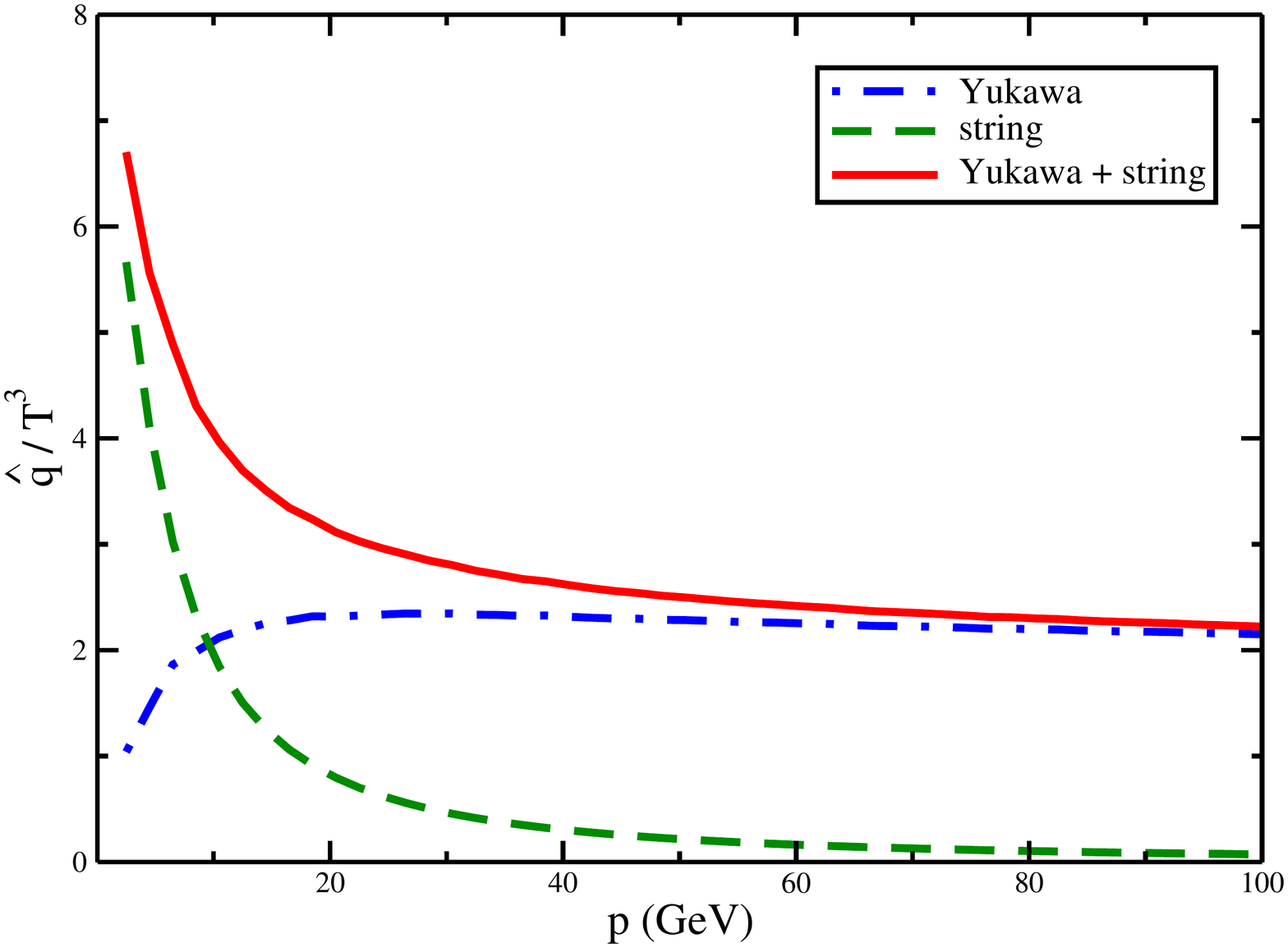}
\caption{(Color online) Upper: The temperature dependence of $\hat{q}/T^3$ for charm quarks with 10~GeV momentum, compared to $\hat{q}/T^3$ for 10~GeV light quarks extracted by the JET collaboration~\cite{JET:2013cls}. Lower: The momentum dependence of $\hat{q}/T^3$ for charm quarks in a QGP medium with 300~MeV temperature.}
\label{Fig:HQ-qhat_over_T3}
\end{figure}

We first present the $R_\mathrm{AA}$ and $v_2$ of $D$ mesons in Pb+Pb collisions at $\sqrt{s_\mathrm{NN}}=5.02$~TeV in Fig.~\ref{Fig:LHC-Dmeson_RAA_v2}, compared to the CMS~\cite{CMS:2017qjw,CMS:2017vhp} and ALICE~\cite{ALICE:2018lyv,ALICE:2017pbx} data. In the figure, we investigate the contributions from the short-range Yukawa and long-range string interactions to the final $D$ meson observables. One can clearly observe that the string interaction dominates the heavy quark energy loss at low $p_\mathrm{T}$ while the Yukawa interaction dominates at high $p_\mathrm{T}$. Since the string interaction has a strong enhancement when the temperature decreases towards the transition temperature $T_\mathrm{c}$ (as will be illustrated in Fig.~\ref{Fig:HQ-qhat_over_T3}), the inclusion of the string interaction helps heavy quarks to obtain larger $v_2$ from the later stage of the QGP which has stronger collectivity.
There are several important factors that may affect the heavy meson $v_2$: the coupling to the collectively expanding medium, the energy loss through the geometrically asymmetric medium, and the coalescence of partons in the hadronization process.
The coupling to the medium collectivity dominates at low $p_\mathrm{T}$, the energy loss dominates at high $p_\mathrm{T}$, and the coalescence mainly affects heavy flavor $v_2$ at low and medium $p_\mathrm{T}$ regions~\cite{Li:2020kax}.
By combining Yukawa and string interactions between heavy quarks and the QGP, our improved LBT model provides a reasonable description of the $D$ meson $R_\mathrm{AA}$ and $v_2$ at the LHC.

Figure~\ref{Fig:RHIC-Dmeson_RAA_v2} shows $R_\mathrm{AA}$ and $v_2$ of $D$ mesons in Au+Au collisions at $\sqrt{s_\mathrm{NN}}=200$~GeV, compared to the STAR data~\cite{STAR:2014wif,STAR:2018zdy,STAR:2017kkh}. We observe that in the $p_\mathrm{T}$ regime concentrated by the STAR measurement, $D$ meson observables are almost entirely dominated by the string interaction. In contrast, the Yukawa interaction alone significantly overestimates the $D$ meson $R_\mathrm{AA}$ and underestimates the $v_2$. Therefore, we expect that the LHC data at high $p_\mathrm{T}$ can help to constrain the Yukawa interaction, while the RHIC data can strongly constrain the string part of the potential between heavy quarks and the QGP.

\begin{figure}[tbp]
\centering
\includegraphics[width=0.97\linewidth]{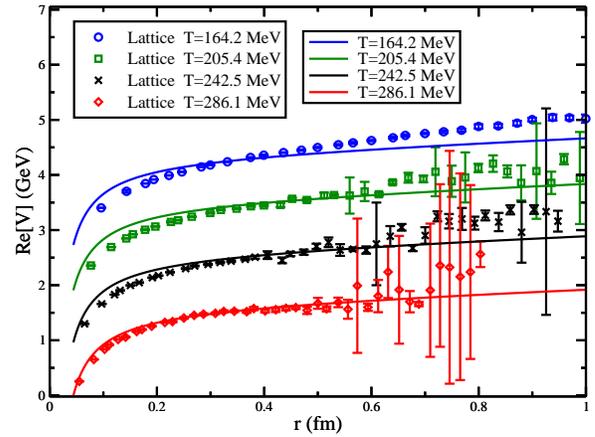}
\caption{(Color online) The in-medium heavy quark potential extracted from our LBT model, compared to the lattice QCD data~\cite{Burnier:2014ssa}.}
\label{Fig:HQ-potential}
\end{figure}

With the model parameters ($\alpha_\mathrm{s}$, $\sigma$, $m_d$ and $m_s$) determined by comparing our heavy flavor $R_\mathrm{AA}$ and $v_2$ results to the experimental data, we calculate the in-medium heavy quark potential according to Eq.~(\ref{eq:Cornell_poten1}) at different temperatures. The result is presented in Fig.~\ref{Fig:HQ-potential} and compared to the lattice QCD data~\cite{Burnier:2014ssa} for the real part of the static quark-antiquark potential in a (2+1)-flavor QCD medium.
Note that there may exist a constant term (independent of $r$) on the right hand side of Eq.~(\ref{eq:Cornell_poten1}).
Such constant term does not generate interaction force between heavy quarks and the QGP, thus cannot be constrained by our transport model calculation.
To compare with the lattice QCD data, we add a constant 5.4~GeV to our results from Eq.~(\ref{eq:Cornell_poten1}).
From Fig.~\ref{Fig:HQ-potential}, one can observe a $-1/r$ dependence at small $r$ and a linear dependence on $r$ at large $r$, which correspond to the Yukawa and string parts of the potential, respectively.
At low temperature (around $T_\mathrm{c}$), the lattice QCD data show a strong confining interaction.
At high temperature (286~MeV and above), the string interaction is significantly screened, leading to a flatter $r$-dependence of the potential at large $r$.
The in-medium potential extracted from our improved LBT model calculation agrees well with the lattice QCD data at high temperature, but show some deviation at low temperature.
One possible reason is the simplified parametrization of the string term in the current study, which limits the temperature and distance (or momentum) dependences of the extracted potential.
Systematic extraction of the in-medium heavy quark potential using more generalized parametrization and the advanced Bayesian statistical analysis method will be carried out in a follow-up study.

\begin{figure}[tbp]
\centering
\includegraphics[width=0.97\linewidth]{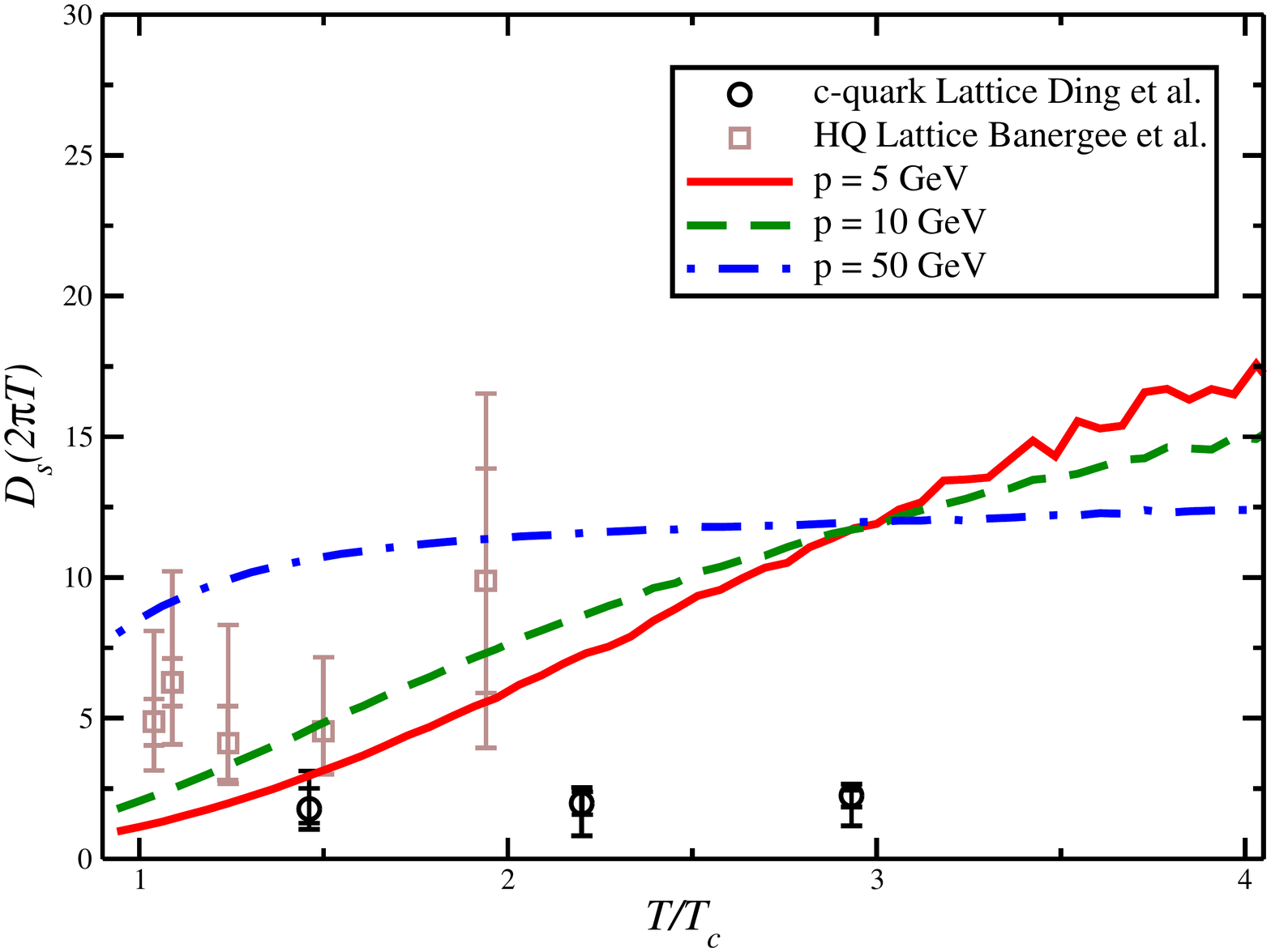}
\includegraphics[width=0.97\linewidth]{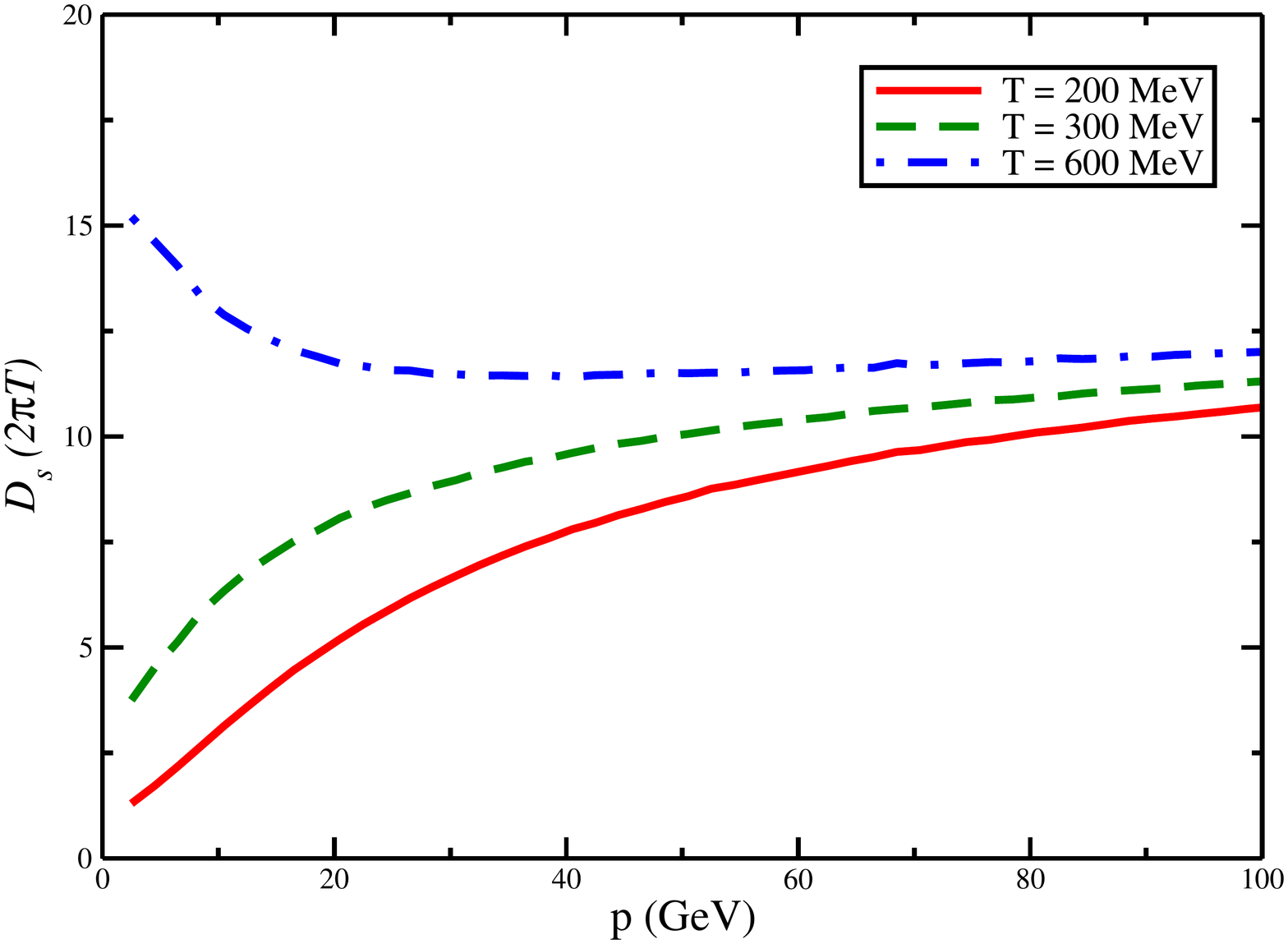}
\caption{(Color online) Upper: The $D_\mathrm{s}(2\pi T)$ as a function of the medium temperature for charm quarks with momenta 5, 10 and 50~GeV, compared to the lattice QCD data~\cite{Banerjee:2011ra, Ding:2012sp}. Lower: The $D_\mathrm{s}(2\pi T)$ as a function of the charm quark momentum for medium temperatures 200, 300 and 600~MeV.}
\label{Fig:HQ-D2piT}
\end{figure}

To further quantify the interaction strength between heavy quarks and the QGP, we calculate the transport coefficients of heavy quarks. In Fig.~\ref{Fig:HQ-qhat_over_T3}, we present the temperature-scaled jet quenching parameter $\hat{q}/T^3$.
The upper panel shows the temperature dependence of $\hat{q}/T^3$ for heavy quarks with 10~GeV momentum, and the lower panel shows the momentum dependence of $\hat{q}/T^3$ for heavy quarks in the medium of temperature 300~MeV. As discussed in Sec.~\ref{sec:new_M2}, the transport parameter $\hat{q}$ can be evaluated using Eq.~(\ref{eq:gamma_el}) weighted by the transverse momentum broadening ($k_\perp^2$) of a given heavy quark. In the upper panel of Fig.~\ref{Fig:HQ-qhat_over_T3}, one observes a much stronger temperature dependence of $\hat{q}/T^3$ from the string interaction than from the Yukawa interaction. The string interaction is strong around $T_\mathrm{c}$ but vanishes at high temperature. After combining the Yukawa and string interactions, our heavy quark $\hat{q}/T^3$ is comparable to the value for 10~GeV light quarks extracted by the JET collaboration work~\cite{JET:2013cls}.
In the lower panel of Fig.~\ref{Fig:HQ-qhat_over_T3}, we also find very different momentum dependences of $\hat{q}/T^3$ between Yukawa and string contributions. For a QGP medium with 300~MeV, $\hat{q}/T^3$ caused by the Yukawa interaction first increases then becomes flat as a function of heavy quark momentum, whereas $\hat{q}/T^3$ from the string interaction decreases with increasing heavy quark momentum. 
In summary, we can clearly see from Fig.~\ref{Fig:HQ-qhat_over_T3} that the heavy-quark-QGP interaction is dominated by the Yukawa term at high temperature and large momentum, but dominated by the string term at low temperature and small momentum.

In Fig.~\ref{Fig:HQ-D2piT}, we show the spatial diffusion coefficient of charm quarks $D_\mathrm{s}(2\pi T)$ after combining the Yukawa and string interactions. Making use of the expression $D_\mathrm{s} = \kappa/(2M^2 \eta_\mathrm{D}^2)$, where $\eta_\mathrm{D}$ is the longitudinal drag and $\kappa$ is the momentum space diffusion coefficient ($\hat{q}=2\kappa$), and the fluctuation-dissipation relation $\eta_\mathrm{D} = \kappa / (2TE)$, one may convert $\hat{q}$ into $D_\mathrm{s}$ via $D_\mathrm{s}(2\pi T)=8\pi/(\hat{q}/T^3)$.
The upper panel shows the temperature dependence of $D_\mathrm{s}(2\pi T)$ for heavy quarks with three different momenta (5, 10 and 50~GeV).
The lower panel shows the momentum dependence of $D_\mathrm{s}(2\pi T)$ for heavy quarks in the QGP medium with three different temperatures (200, 300 and 600~MeV).
From the upper panel of Fig.~\ref{Fig:HQ-D2piT}, we can see that our extracted $D_\mathrm{s}(2\pi T)$ increases as a function of the medium temperature.
The slope of increase is smaller for heavy quarks with larger momenta due to the increasing contribution from the Yukawa part of the interaction.
Our results for $D_\mathrm{s}(2\pi T)$ are comparable to the lattice QCD data from Ref.~\cite{Banerjee:2011ra} and~\cite{Ding:2012sp}.
In the lower panel of Fig.~\ref{Fig:HQ-D2piT}, we can see that at low temperature, $D_\mathrm{s}(2\pi T)$ increases with increasing heavy quark momentum due to the dominant contribution from non-perturbative string interaction, while at high temperature, it first decreases then becomes flat as a function of the heavy quark momentum because of the increasing contribution from the Yukawa interaction.

\section{Summary}
\label{sec:summary}

We have developed a unified approach for simulating both perturbative and non-perturbative interactions between heavy quarks and the QGP in relativistic heavy-ion collisions. In our new approach, the propagator of the exchanged gluon between a heavy quark and a thermal parton from the medium is replaced by a general Cornell-type potential that incorporates both short-range Yukawa and long-range string interactions. Using the generalized potential, we have computed the scattering matrix elements and then implemented them in the linear Boltzmann Transport (LBT) model. By coupling the improved LBT model to a (3+1)-D viscous hydrodynamic model CLVisc and a hybrid fragmentation-coalescence model, we have performed a realistic simulation of heavy quark in-medium evolution and hadronization in high-energy nuclear collisions.

Using our improved LBT model, we find that heavy quark observables at high $p_\mathrm{T}$ are dominated by the short-range Yukawa potential, while at low $p_\mathrm{T}$ they are dominated by the long-range string potential.
Since the string interaction increases rapidly as the medium temperature decreases towards the transition temperature $T_\mathrm{c}$, the inclusion of the non-perturbative string term enhances the heavy quark elliptic flow $v_2$ due to their later stage interaction with the QGP.
By combining the perturbative and non-perturbative contributions, our improved LBT model is able to consistently describe the $D$ meson $R_\mathrm{AA}$ and $v_2$ at both RHIC and the LHC.
Through the model-to-data comparison, we have made the first extraction of the in-medium heavy quark potential from the open heavy flavor observables.
Our extracted potential is in reasonable agreement with the lattice QCD data.
We have also computed the heavy quark transport coefficients, including the jet quenching parameter $\hat{q}/T^3$ and the spatial diffusion coefficient $D_\mathrm{s} (2\pi T)$ as functions of the medium temperature and the heavy quark momentum.
Our results on the heavy quark in-medium potential and transport coefficients show that while the short-range Yukawa potential dominates heavy-quark-QGP interaction at high temperature and large momentum, the long-range color confining potential dominates at low temperature and small momentum.

Our work constitutes a more quantitative understanding of heavy quark dynamics inside a color de-confined nuclear medium. In particular, we have provided a proof-of-principle extraction of the in-medium heavy quark potential using the open heavy flavor observables. More precise determination of the interaction potential between heavy quarks and the QGP using a more general ansatz of its temperature and distance dependences, together with a more systematic statistical analysis routine such as the Bayesian method, will be implemented in a follow-up effort.

\section*{Acknowledgments}

We are grateful to helpful discussions with Shuai Liu and Xiang-Yu Wu. This work was supported by the National Natural Science Foundation of China (NSFC) under Grant Nos. 11775095, 11890710, 11890711, 11935007, 12175122 and 2021-867. W.-J.X. would like to thank the support from the CCNU Huabo Program. Some of the calculations were performed in the Nuclear Science Computing Center at Central China Normal University (NSC$^3$), Wuhan, Hubei, China.

\bibliographystyle{h-physrev5}
\bibliography{SCrefs}

\end{document}